\global\def\draftcontrol{0}

   \def\versionno{ n4 bh instability}
\catcode`\@=11

\expandafter\ifx\csname draftcontrol\endcsname\relax\global\def\draftcontrol{0}
\fi

{\count255=\time\divide\count255 by 60
\xdef\hourmin{\number\count255}
\multiply\count255 by-60\advance\count255 by\time
\xdef\hourmin{\hourmin:\ifnum\count255<10 0\fi\the\count255}}
\def\draftdate{\number\month/\number\day/\number\year\ \ \ \hourmin }

\newcommand\makepapertitle{\par
  \begingroup
    \renewcommand\thefootnote{\@fnsymbol\c@footnote}%
    \def\@makefnmark{\rlap{\@textsuperscript{\normalfont\@thefnmark}}}%
    \long\def\@makefntext##1{\parindent 1em\noindent
            \hb@xt@1.8em{%
                \hss\@textsuperscript{\normalfont\@thefnmark}}##1}%
     \newpage
     \global\@topnum\z@   % Prevents figures from going at top of page.
     \@makepapertitle
     \thispagestyle{empty}\@thanks
  \endgroup
  \setcounter{footnote}{0}%
  \global\let\thanks\relax
  \global\let\makepapertitle\relax
  \global\let\@makepapertitle\relax
  \global\let\@thanks\@empty
  \global\let\@author\@empty
  \global\let\@date\@empty
  \global\let\@title\@empty
  \global\let\title\relax
  \global\let\author\relax
  \global\let\date\relax
  \global\let\and\relax
  \def\version{\let\version\@version\@gobble}
}
\def\@makepapertitle{%
  \newpage
   \ifnum\draftcontrol=1 {}
   \version\versionno
   \vskip 3em%
   \else
   \hfill\hbox to 3cm {\parbox{4cm}{\@pubnum}\hss}%
   \vskip 3em%
   \fi
   \begin{center}%
   \let \footnote \thanks
     {\LARGE {\@title}}%
     \vskip 1.5em%
     {\normalsize%\large
       \lineskip .5em%
       \begin{tabular}[t]{c}%
         \@author
       \end{tabular}\par}%
     \vskip 1.5em%
     {\@bstract}%
     \end{center}%
     \vskip 1.5em
     \@date%
   \par
}

\gdef\@pubnum{}
\def\pubnum#1{%
  \gdef\@pubnum{#1}}

\gdef\@bstract{}
\def\Abstract#1{%
  \gdef\@bstract{%
   \parbox{\textwidth-0pc}{%
   \centerline{\bf Abstract}\penalty1000%
\kern.2cm%
\noindent%\abstractfont \baselineskip=12pt
\renewcommand\baselinestretch{1.0}%
{#1}}}
}

\def\ps@paper{\let\@mkboth\@gobbletwo%
     \ifnum\draftcontrol=1
    \def\@oddfoot{\hbox to \textwidth{\tiny \versionno \hfil\tiny\draftdate}%
    \hskip -\textwidth \hbox to \textwidth{\hfil\rm\thepage\hfil}}%
     \else\def\@oddfoot{\hbox to \textwidth{\hfil\rm\thepage\hfil}}
     \fi
     \let\@evenfoot\@oddfoot
}
\def\body{\clearpage
          \pagestyle{paper}
    }
\def\@version#1{\ifnum\draftcontrol=1
\typeout{}\typeout{#1}\typeout{}
\vskip3mm\centerline{\hbox{\fbox{\normalsize{\tt DRAFT -- #1 -- }
                   {\draftdate}}}}\vskip3mm
\fi}
\let\version\@version
\long\def\eqlabel#1{\ifnum\draftcontrol=1
                    \tag@false  % there are some problems with multline without this
                    \tag*{(\theequation) \hbox to -0.2cm{\hspace{0cm}\small{#1}\hss}}
                    \refstepcounter{equation}
                    \edef\@currentlabel{\theequation}
                    \ltx@label{#1}          % use old LaTeX \label instead of new definition
                    \else
                    \label{#1}
                    \fi
                    }
\let\st@bibitem\@bibitem
\let\st@lbibitem\@lbibitem
\ifnum\draftcontrol=1
  \def\@bibitem#1{%
    \st@bibitem{#1}\a@@label{#1}\ignorespaces}
  \def\@lbibitem[#1]#2{%
    \st@lbibitem[#1]{#2}\a@@label{#2}\ignorespaces}
  \def\a@@label#1{%
    \gdef\a@lab{\smash{\normalfont\small#1}}
    \ifvmode
      \if@inlabel
        \global\setbox\@labels\hbox{%
          \llap{\a@lab\let\a@lab\relax
                \kern\@totalleftmargin\kern\marginparsep}%
          \box\@labels}%
      \fi
    \fi}
\fi
\documentclass[12pt,letterpaper]{article}
\usepackage{amsmath,amssymb,array,calc,epsfig,rotating,bm,xcolor}
\usepackage[sort]{cite}
\usepackage{graphicx,esint,float}
\usepackage{psfrag,verbatim}
\usepackage[makeroom]{cancel}
\usepackage{xcolor,url}
\usepackage{hyperref}
\ifnum\draftcontrol=1
\fi
\renewcommand\baselinestretch{1.25}
\renewcommand\section{\@startsection {section}{1}{\z@}%
                                   {-3.5ex \@plus -1ex \@minus -.2ex}%
                                   {2.3ex \@plus.2ex}%
                                   {\normalfont\large\bfseries}}
\renewcommand\subsection{\@startsection{subsection}{2}{\z@}%
                                   {-3.25ex\@plus -1ex \@minus -.2ex}%
                                   {1.5ex \@plus .2ex}%
                                   {\normalfont\normalsize\bfseries}}
\renewcommand\subsubsection{\@startsection{subsubsection}{3}{\z@}%
                                   {-3.25ex\@plus -1ex \@minus -.2ex}%
                                   {1.5ex \@plus .2ex}%
                                   {\normalfont\normalsize\it}}
\renewcommand\paragraph{\@startsection{paragraph}{4}{\z@}%
                                   {-3.25ex\@plus -1ex \@minus -.2ex}%
                                   {1.5ex \@plus .2ex}%
                                   {\normalfont\normalsize\bf}}

\numberwithin{equation}{section}

\def\revise#1       {\raisebox{-0em}{\rule{3pt}{1em}}%
                     \marginpar{\raisebox{.5em}{\vrule width3pt\
                     \vrule width0pt height 0pt depth0.5em
                     \hbox to 0cm{\hspace{0cm}{%
                     \parbox[t]{4em}{\raggedright\footnotesize{#1}}}\hss}}}}

\newcommand{\ie}{{\it i.e.,}\ }
\newcommand{\eg}{{\it e.g.,}\ }

\def\cald         {{\cal D}}
\def\cale         {{\cal E}}

\def\calh         {{\cal H}}

\def\calm         {{\cal M}}
\def\caln         {{\cal N}}
\def\calo         {{\cal O}}

\def\reals        {{\mathbb R}}
\def\zet          {{\mathbb Z}}

\def\del          {\partial}

\def\Im           {{\rm Im\hskip0.1em}}

\def\sqr#1#2{{\vcenter{\vbox{\hrule height.#2pt
 \hbox{\vrule width.#2pt height#1pt \kern#1pt
 \vrule width.#2pt}\hrule height.#2pt}}}}
\def\square{%
  \mathop{\mathchoice{\sqr{12}{15}}{\sqr{9}{12}}{\sqr{6.3}{9}}{\sqr{4.5}{9}}}}

\newcommand{\kk}{\mathfrak{q}}
\newcommand{\ww}{\mathfrak{w}}

\def\kt{\tilde{k}}
\def\a{\alpha}

\def\aa1{\phi}
\def\cc1{\psi}

\def\tk{\tilde{k}}

\def\f0{\text{\boldmath$\varphi$}}
\def\h2{\mathfrak{h}}
\def\at{\mathfrak{a}_t}
\def\ar{\mathfrak{a}_r}
\def\az{\mathfrak{a}_z}

\catcode`\@=12

\begin{document}

%%%
%%%%%% text starts here
%%%%%%%%%

\title{\bf Instability in ${\cal N}=4$ supersymmetric
Yang-Mills theory on $S^3$ at finite density}

\date{June 21, 2026}
%\date\today

\author{
Alex Buchel\\[0.4cm]
\it Department of Physics and Astronomy\\ 
\it University of Western Ontario\\
\it London, Ontario N6A 5B7, Canada
}

\Abstract{Homogeneous and isotropic equilibrium states of strongly coupled
${\cal N}=4$ supersymmetric Yang-Mills charged plasma in ${\mathbb
R}^3$ with equal chemical potentials for the maximal Abelian subgroup
of the $R$-symmetry group become dynamically unstable below some
critical temperature.  The instabilities arise in the $R$-symmetry
charge transport, precisely when the equilibrium state becomes
thermodynamically unstable.  We study the fate these correlated
instabilities when the theory is placed on $S^3$. The curvature of the
three-sphere affects the onset of the dynamical and the thermodynamic
instabilities differently: increasing the curvature at low
temperatures can stabilize its transport, but leave the plasma
thermodynamically unstable.  Thermodynamic stability is never recovered
provided the $S^3$ volume is allowed to fluctuate. $\caln=4$ SYM is
always thermodynamically unstable in the fixed-pressure ensemble
on $S^3$ as well. 
}

\makepapertitle

\body

\version\versionno
\tableofcontents

\section{Introduction and summary}\label{intro}

Hydrodynamics is a universal description of small fluctuations
about the thermal equilibrium states of a relativistic plasma.
Naturally, the hydrodynamic transport coefficients
are sensitive to the thermodynamics of an equilibrium:
\eg the relativistic neutral plasma at temperature $T$
in $\reals^3$ propagates
energy-pressure fluctuations as sound waves with the
dispersion relation\footnote{The linearized profile for the
fluctuations is assumed to be proportional $\exp(-i w t +i \bm{k}
\bm{x})$.} 
\begin{equation}
\ww(\kk)=\pm c_s \kk +\calo(\kk^2)\,,\qquad \ww\equiv \frac{w}{2\pi T}\,,
\qquad \kk\equiv \frac{|\bm{k}|}{2\pi T}\,.
\eqlabel{sound}
\end{equation}
The speed of the sound waves, $c_s$, is determined by the equilibrium
equation of state
\begin{equation}
c_s^2=\frac{\del P}{\del\cale}=\frac{s}{c_V}\,,
\eqlabel{cs}
\end{equation}
where $P,\cale$ are the equilibrium  pressure and
energy density; we further used the thermodynamic relations
to express the sound speed in terms of the entropy density $s$
and the specific heat $c_V$. Whenever the equilibrium state
is thermodynamically unstable, \ie $c_V<0$, the speed of the sound
waves becomes imaginary, and such a homogeneous and isotropic
equilibrium state of the plasma becomes unstable to energy density
clumping. At strong coupling, a large class of
quantum gauge theories  have dual holographic descriptions as
classical gravitational theories in asymptotically $AdS_5$ spacetime
\cite{Maldacena:1997re,Aharony:1999ti}. The thermal states
of the gauge theory plasma are mapped \cite{Witten:1998zw} to  
black branes\footnote{Or black holes if  the boundary gauge theory
is formulated on a compact space.} of the gravitational dual. 
The precise correlation between the thermodynamic instability
of the plasma and its hydrodynamic instability directly
translates into the correlated
instabilities\footnote{Correlated (in)stability conjectures
were originally discussed in \cite{Gubser:2000mm,Gubser:2000ec}.}
of the dual black
branes\footnote{See \cite{Buchel:2018bzp,Buchel:2020jfs}
for additional holographic examples.} \cite{Buchel:2005nt}.

In \cite{Gladden:2024ssb,Gladden:2025glw} the authors (GIKS)
extended the argument of  \cite{Buchel:2005nt} to holographic examples
of the boundary gauge theory charged plasma equilibrium states.
There, the thermodynamic instability identified in the Hessian
matrix of the  plasma equation of state $\cale=\cale(s,\rho_\alpha)$,
where $\rho_\alpha$ are various equilibrium state charge 
densities, leads to the charge transport with the negative
diffusion coefficient, resulting in charge density (as
opposed to energy density in the neutral plasma) clumping.
GIKS focused on the equilibrium states of the
strongly coupled $\caln=4$ supersymmetric
Yang-Mills (SYM) plasma with the same\footnote{We refer to such states
as {\it symmetric}.} chemical potentials $\mu_\alpha\equiv \mu$
for $U(1)^3\subset SU(4)$ $R$-symmetry. Both the thermodynamic
and the correlated hydrodynamic instability develops when
the temperature drops below the
critical value $T<T_{crit}$,
\begin{equation}
\frac{\mu}{2\pi T_{crit}}=\sqrt 2\,.
\eqlabel{mutc}
\end{equation}
This instability precludes the existence of  homogeneous and
isotropic symmetric charged states of $\caln=4$ plasma
in the extremal limit $T\to 0$ with finite entropy density,
as described by AdS-Reissner-Nordstr\"om black
branes\footnote{See \cite{Buchel:2025jup,Buchel:2025ven,Buchel:2026nlg}
for additional examples
of GIKS instability in the holographic models.}.
It sets in at a higher temperature than the familiar 
superconducting instability \cite{Buchel:2025ves}.  

In this work we extend the analysis of \cite{Gladden:2024ssb}
to symmetric charged states of the $\caln=4$ SYM plasma on a
three-sphere. The curvature of the $S^3$,
\begin{equation}
K\equiv \frac{1}{R_{S^3}^2}\,,
\eqlabel{defcurv}
\end{equation}
modifies both the equilibrium thermodynamics of the theory, and
the spectrum of charged plasma excitations --- the quasinormal modes
(QNMs) --- of the corresponding gravitational dual black holes.
In Appendix \ref{eomshol} we extend the Kodama-Ishibashi
{\it master field formalism} for QNM analysis
\cite{Kodama:2003jz,Jansen:2019wag,Buchel:2021ead}
to gravitational models with arbitrary number of scalar and gauge fields,
along with generic couplings and the scalar potential. 
The developed formalism is further applied, see Appendix \ref{stuqnm},
to gravitational
STU models \cite{Behrndt:1998jd,Cvetic:1999xp,Harmark:1999xt,Cvetic:1999ne}
realizing the holographic duals of charged $\caln=4$ SYM
plasma states at strong coupling.

\begin{figure}[t]
\begin{center}
\psfrag{d}{{$\cald$}}
\psfrag{x}[r]{{$K/(\pi T_0)^2$}}
\psfrag{y}[t]{{$\Im[\ww^{(\ell)}]$}}
\psfrag{k}{{$\kappa$}}
\psfrag{a}[r]{{$\ell=1$}}
 \psfrag{b}[bl]{{$\ell=2$}}
 \psfrag{c}{{$\ell=3$}}
  \includegraphics[width=3in]{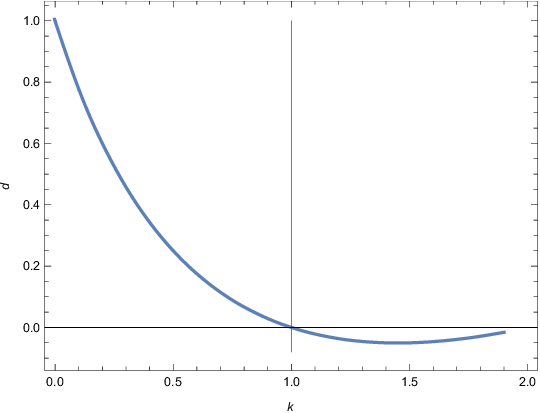}
\ \includegraphics[width=3in]{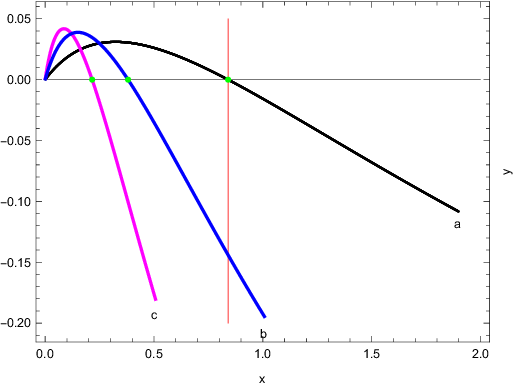}
\end{center}
 \caption{The left panel: we use the master field formalism
 of Appendix \ref{eomshol} to reproduce the diffusion coefficient
 $\cald$ \eqref{defcald} of
 the $\caln=4$ SYM plasma with equal chemical potentials in $\reals^3$
reported in \cite{Gladden:2024ssb}. The right panel: we study diffusive
modes in the charged SYM plasma on $S^3$ of radius $\frac{1}{\sqrt K}$.
Different $\ell$ diffusive QNMs \eqref{defell} are stabilized
at increasing values of $K=K_{crit}^{(\ell)}$, represented by the green
dots. The last mode to stabilize is $\ell=1$.
}\label{figure1}
\end{figure}

We now summarize our findings.
In the left panel of fig.~\ref{figure1} we reproduce the
results\footnote{We independently verified the computation
using the ref.~\cite{Gladden:2024ssb} equations of motion for the
QNMs.} of
\cite{Gladden:2024ssb} for the
$R$-charge diffusion coefficient $\cald$ using the formalism of
Appendix \ref{eomshol}. Here, the helicity $h=0$ channel dispersion relation
for the diffusive modes takes the form 
\begin{equation}
\ww_{diff}=-i \cald\ \kk^2+\calo(\kk^4)\,,
\eqlabel{defcald}
\end{equation}
and  $\kappa\in [0,2]$ parameterizes the ratio $\frac{T}{\mu}$ for the
symmetric equilibrium state of the SYM plasma, see \eqref{stuthermo},
\begin{equation}
\frac{2\pi T}{\mu} = \frac{2-\kappa}{\sqrt{2\kappa}}\,.
\eqlabel{symtmu}
\end{equation}
The diffusion coefficient $\cald$ vanishes at $\kappa=\kappa_{crit}=1$
(highlighted by the red line),
and becomes negative for $\kappa> 1$, correspondingly for
$T<T_{crit}$ in \eqref{mutc}. Once the SYM is placed on $S^3$,
the fluctuations carry a spherical harmonics index $\ell$,
rather than a continuous label $\bm{k}$, related
to the eigenvalue of the Laplacian on $S^3$, see \eqref{ks3},
\begin{equation}
\bm{k}^2\ \Longrightarrow\ k^2\equiv K\ell (\ell+2)\,,\qquad \ell=0,1,2\cdots\,,
\eqlabel{defell}
\end{equation}
where $K$ is defined as in \eqref{defcurv}.
The QNM frequency $\ww$ depends both on a discrete label
$\ell$ and on a continuous parameter $\frac{K}{T_0^2}$,
where $T_0$ is an arbitrary auxiliary scale characterizing the
thermal state of the $\caln=4$ SYM on $S^3$, see \eqref{stuthermo}. 
In the right panel of fig.~\ref{figure1} we set $\kappa=\frac32$
and analyze the diffusive QNMs. While there is no instability in the $\ell=0$
modes\footnote{This is not even a diffusive mode as the
gauge fields fluctuations are not dynamical in this case,
see Appendix \ref{h0l0s}.}, all higher-$\ell$ modes are unstable
in the limit $K/T_0^2\to 0$ --- we show this explicitly
for $\ell=1,2,3$ QNMs
(the black, blue, and magenta curves, respectively).
As $K/T_0^2$ increases, the higher-$\ell$
modes are stabilized first, $\Im[\ww]<0$, crossing the $K$-axis
at green dots, located at $K_{crit}^{(\ell)}$,
\begin{equation}
\Im\left[\ww^{(\ell)}(K)\right]\bigg|_{K=K_{crit}^{(\ell)}}=0\,,\qquad
{\rm with}\qquad 
K_{crit}^{(\ell')}<K_{crit}^{(\ell)}\ \ \  {\rm if}\ \ \  \ell'> \ell \,.
\eqlabel{defkcriti}
\end{equation}
The last mode to stabilize is $\ell=1$, which determines
the critical value $K_c=K_{crit}^{(1)}$
for the onset of the dynamical instability
in the SYM plasma: the symmetric equilibrium
charged state of the $\caln=4$ SYM plasma on $S^3$
is dynamically stable provided $K>K_c$
for a given temperature $T$ and chemical potential(s) $\mu_\alpha=\mu$,
see \eqref{stuthermo}.

\begin{figure}[t]
\begin{center}
\psfrag{k}{{$\kappa-1$}}
\psfrag{s}[b]{{$K_{crit}^{(\ell)}/(\pi T_0)^2$}}
\psfrag{g}[t]{{$K_{c}/(\pi T_0)^2$}}
  \includegraphics[width=3in]{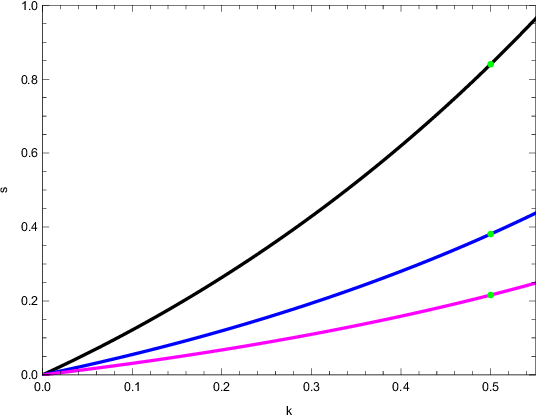}
\  \includegraphics[width=3in]{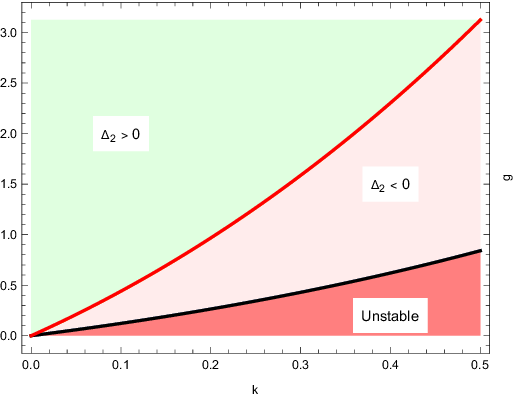}
\end{center}
 \caption{The left panel: the critical values $K_{crit}^{(\ell)}$
required to stabilize the diffusive QNMs,
$\Im[\ww^{(\ell)}(K>K_{crit}^{(\ell)})]<0$, as a function
of $\kappa$ \eqref{stuthermo}. The right panel:
charged  $\caln=4$ SYM plasma on $S^3$ of curvature $K$ is thermodynamically
unstable from the criteria $\Delta_2>0$ (see \eqref{eq4c1}
) below the red curve, and is dynamically unstable
below the black curve. 
}\label{figure2}
\end{figure}

In the left panel of fig.~\ref{figure2} we plot $K_{crit}^{(\ell)}$
for $\ell=1,2,3$ as a function of $\kappa$, see \eqref{stuthermo}.
We use the same color coding as in the right panel of
 fig.~\ref{figure1}. In the right panel of fig.~\ref{figure2}
 we plot $K_c$ for the onset of the dynamical instability
 in the charged plasma, the black curve, and the
 onset of its thermodynamic instability from the $\Delta_2$
 leading minor of the plasma equation of state $E(Q_\alpha,S,V)$ Hessian,
 represented by the red curve,
 see Appendix \ref{stuthermos} and in particular \eqref{eq4c2}.
The $\Delta_2<0$  criteria at $K=0$ in \cite{Gladden:2024ssb}
identified $\kappa>1$ as thermodynamically unstable region of the
symmetrically ($\mu_\alpha\equiv  \mu$) charged
$\caln=4$ SYM plasma. For $K>0$, the charged SYM plasma
 is  dynamically stable and $\Delta_2>0$ in the
 light-green region in the right panel of fig.~\ref{figure2};
 it is dynamically unstable and has $\Delta_2<0$ in the
 red region; finally, in between the black and the red curves
 (the light-pink region) the equilibrium states of the
 charged SYM plasma on $S^3$ are dynamically stable, but are
 thermodynamically unstable due to $\Delta_2<0$.
 As emphasized in Appendix \ref{stuthermos}, if the volume
 of $S^3$ is allowed to fluctuate, the 
 spatially homogeneous and isotropic equilibrium states of
 charged $\caln=4$ SYM are never stable
 (at least for small $K$); the equilibrium states of
 neutral ($\mu_\alpha= 0$) $\caln=4$ SYM plasma on $S^3$
 are thermodynamically unstable for any  curvature $K>0$. 
In Appendix \ref{stuthermos2} we analyze
the thermodynamic stability of $\caln=4$ SYM plasma on $S^3$
in the enthalpy representation (fixed pressure) ensemble:
the conclusion is identical to the discussion in the microcanonical
ensemble of Appendix \ref{stuthermos}.

In this paper we analyzed the dynamical and thermodynamic instabilities of
the $\caln=4$ SYM on $S^3$. Contrary to the perfect correlation
between these instabilities in the three-sphere decompactification limit
$S^3\to \reals^3$, here we encounter states that are dynamically
stable, but are nonetheless thermodynamically unstable.
We believe this is the first reported occurrence of this phenomenon ---
in \cite{Buchel:2020jfs} a different breakdown of the
correlated stability was reported: there, the thermal equilibrium
states of the model were thermodynamically stable, but dynamically
unstable.

It is an interesting open question whether there is a new preferred
phase of $\caln=4$ SYM on $S^3$ when it is dynamically stable,
but thermodynamically unstable. The absence of dynamical instabilities suggests that if there is
a new thermodynamically stable phase,
it is not continuously connected to a homogeneous and isotropic
phase of $\caln=4$ plasma on $S^3$. Another open question is
the origin of the thermodynamic instability
of $\caln=4$ SYM on $S^3$ either in the
microcanonical ensemble with $dV\ne 0$, or the enthalpy representation ensemble.
We hope to report on these
issues in the future.

\section*{Acknowledgments}
I would like to thank Pavel Kovtun for valuable discussions.
This work was supported by NSERC through the Discovery Grants program.

\appendix
\section{Master equations of Einstein-Maxwell-scalar black holes in $D=5$}\label{eomshol}
In this section we extend the work of \cite{Buchel:2021ead} to black holes
in theories of Einstein gravity in $D=5$ space-time dimensions with multiple
scalars and multiple gauge fields. This extends the formalism of \cite{Jansen:2019wag}
where a single scalar field and a single bulk gauge field were considered. 
We closely follow the notations of \cite{Buchel:2021ead}.

Consider an effective action
\begin{equation}
S_5=\int_{\calm_5}d^{3+2}\xi\ \sqrt{-g} \biggl[R-\sum_{j=1}^p \eta_j \left(\del\phi_j\right)^2
-\frac 14 \sum_{\alpha=1}^q Z_\alpha(\{\phi_i\})\ F_\alpha^2-V\left(\{\phi_i\}\right)\biggr]\,,
\eqlabel{efac}
\end{equation}
where $j=1\cdots p$ indexes the scalars $\phi_i$; $\eta_i$'s are the constant normalizations of the scalar
kinetic terms;
$\alpha=1\cdots q$ indexes the gauge fields $F_\alpha=dA_\alpha$, 
and $Z$ and $V$ are arbitrary functions of the scalar fields.
We will be interested in the stability analysis of the
black branes/holes in the theory \eqref{efac} with maximally symmetric 3-dimensional Schwarzschild horizons:
\begin{equation}
ds_5^2=-c_1^2\ dt^2+ c_2^2\ dX_{3,K}^2+c_3^2\ dr^2\,,
\eqlabel{5dmetric}
\end{equation}
where $c_i=c_i(r)$, $\phi_j=\phi_j(r)$, $A_\alpha=a_\alpha(r) dt$ and 
\begin{equation}
dX_{3,K}^2=\begin{cases}
d\bm{x}^2\equiv dx_1^2+dx_2^2+dx_3^2\,,\qquad &K=0\,,\qquad {\rm planar}\,,\\
d\Omega_{(3)}^2\,,\qquad  &K>0\,,\qquad {\rm spherical}\,,\\
dH_{(3)}^2\,,\qquad  &K<0\,,\qquad {\rm hyperbolic}\,.
\end{cases}
\eqlabel{defxnk}
\end{equation}
Note that we do not fix $K=\{0,\pm 1\}$, but instead allow it to vary smoothly --- this would allow
for the interpolation of the quasinormal spectra between different maximally symmetric horizons, notably
between the planar and the spherical ones.
A useful way to explicitly parameterize $X_{3,K}$ and a metric on it is as follows:
$X_{3,K}=(x_1\equiv x, x_2\equiv y,  x_3\equiv z)$, and 
\begin{equation}
dX_{3,K}^2=\frac{dx^2}{(1-K x^2)}+(1-K x^2)\ \biggl[\frac{dy^2}{(1-K y^2)}+(1-K y^2)\ dz^2\biggr]\,.
\eqlabel{dx3k}
\end{equation}

From \eqref{efac} we obtain the following second order equations of motion ($'\equiv \frac{d}{dr}$  and
$\del_j\equiv \frac{\del}{\del\phi_j}$)
\begin{equation}
0=c_1''+c_1' \left[\ln\frac{c_2^3}{c_3}\right]'+\frac{c_3^2 c_1}{3}\ V-\frac{1}{3c_1}\sum_\alpha Z_\alpha
\cdot (a'_\alpha)^2\,,
\eqlabel{bac1}
\end{equation}
\begin{equation}
0=c_2''+c_2' \left[\ln\frac{c_1 c_2^2}{c_3}\right]'+\frac{c_2 c_3^2}{3}\ \left(V-\frac{6 K}{c_2^2}\right)+\frac{c_2}{6c_1^2}\sum_{\alpha} Z_a\cdot (a_\alpha')^2\,,
\eqlabel{bac2}
\end{equation}
\begin{equation}
0=\phi_j''+\phi_j' \left[\ln\frac{c_1 c_2^3}{c_3}\right]'-\frac{c_3^2}{2\eta_j}\
\del_jV+\frac{1}{4c_1^2\eta_j}\sum_\alpha \del_jZ_\alpha\cdot (a_\alpha')^2\,,
\eqlabel{bac3}
\end{equation}
\begin{equation}
0=a_\alpha''+a_\alpha' \cdot\biggl\{
\left[\ln\frac{c_2^3}{c_1 c_3}\right]'+[\ln Z_\alpha]'\biggr\}\,,
\eqlabel{bac4}
\end{equation}
and the first order constraint\footnote{We verified that \eqref{bacc} is consistent
with \eqref{bac1}-\eqref{bac4}.}
\begin{equation}
0=\sum_{j=1}\eta_j (\phi_j')^2-\left[\ln c_2^3\right]' \left[\ln(c_1^2 c_2^2)\right]'
+c_3^2\ \left(\frac{6 K}{c_2^2}-V\right)-\frac{1}{2c_1^2} \sum_{\alpha} Z_a\cdot (a_\alpha')^2\,.
\eqlabel{bacc}
\end{equation}

We organize all the gauge invariant fluctuations into three sets of master scalars of different helicity $h$:
\begin{itemize}
\item the helicity $h=2$ set, $\{\Phi_2^{(2)}\}$;
\item the helicity $h=1$ set, $\{\Phi_2^{(1)},\Phi_{(1,\alpha)}^{(1)}\}$, $\alpha=1\cdots q$;
\item the helicity $h=0$ set, $\{\Phi_2^{(0)},\Phi_{(1,\alpha)}^{(0)},\Phi_{(0,j)}^{(0)}\}$, $j=1\cdots p$
and $\alpha=1\cdots q$.
\end{itemize}
Any master scalar $\Phi_s^{(h)}$ ($s=2$, $s=(1,\alpha)$  or $s=(0,j)$ and $(h)=\{0,1,2\}$) is assumed to
have the following dependence:
\begin{equation}
\Phi_s^{(h)}(\xi)=F_{s}^{(h)}(t,r)\ S(X_{3,K})\,,
\eqlabel{masterscalar}
\end{equation}
where $S(X_{3,K})$ is a scalar eigenfunction of the Laplacian $\Delta_K$ on \eqref{dx3k} with an eigenvalue
$k^2$:
\begin{equation}
\Delta_K\ S + k^2\ S=0 \,.
\eqlabel{compact}
\end{equation}
In this work we will be concerned with planar ($K=0$) or  spherical  ($K>0$)  horizons. In the former
case, $k^2 \in [0,+\infty)$ and in the latter case
\begin{equation}
k^2=K \ell (\ell +2)\qquad   {\rm with}\qquad   \ell\in \zet_+\,.
\eqlabel{ks3}
\end{equation}

Each of the master scalars satisfies a coupled master equation of the form
\begin{equation}
\square \Phi_s^{(h)}-W_{s,s'}^{(h)}(r)\ \Phi_{s'}^{(h)}=0\,,
\eqlabel{eqms}
\end{equation}
where $\square$ is the wave operator on the full $D=5$ metric \eqref{5dmetric}, and the symmetric potential
matrix,
\begin{equation}
W_{s,s'}^{(h)}=W_{s',s}^{(h)}\,,
\eqlabel{sym}
\end{equation}
couples master scalars in a given helicity set.

We now present results for potentials $W_{s,s'}^{(h)}$ in different helicity sectors,
as well as relations between the  master scalars and a specific set of gauge invariant
fluctuations in that sector. We refer the reader to \cite{Jansen:2019wag} for a detailed
discussion of exactly how the gauge invariant fluctuations are
constructed\footnote{In a holographic setting gauge invariant fluctuations in Einstein-scalar theories
were used for the first time in \cite{Benincasa:2005iv} and in
Einstein-Maxwell-scalar theories in \cite{Buchel:2010gd}.}.

\subsection{Helicity $h=0$ sector}

The relation between the gauge invariant fluctuations in the scalar sector
\begin{equation}
\{\f0_j\,,\ \at^\alpha\,,\ \ar^\alpha\,,\ \h2_{tr}\,,\ \h2_{rr}\,,\ \h2_{xr}\,,\ \h2_{tt}\}\,,
\eqlabel{gauge0}
\end{equation}
and the master scalars
\begin{equation}
\{F^{(0)}_{(0,j)}\,,\ F^{(0)}_{(1,\alpha)}\,,\ F^{(0)}_2\}\,,
\eqlabel{gauge0m}
\end{equation}
all being functions of $(t,r)$, is as follows:
\begin{equation}
\begin{split}
\f0_j&=-\frac{1}{\sqrt{2\eta_j}}\ F_{(0,j)}^{(0)}+\frac{\sqrt3 k\ c_2\phi_j'}
{6\kt\ c_2'} \ F_2^{(0)} \,,
\end{split}
\eqlabel{h01}
\end{equation}
\begin{equation}
\begin{split}
&\h2_{rr}= \frac{ c_2 c_3^2  (c_2' c_1-c_1' c_2)}{D}\ \sum_{j=1}^p\bigg\{\sqrt{2\eta_i}\phi_j'\
F_{(0,j)}^{(0)}\bigg\}
+\biggl(
\frac{c_2^2 c_3^2\ k}{3\sqrt{3}(c_2')^2\ \kt}\ \sum_{j=1}^p \bigg\{\eta_j (\phi_j')^2\bigg\}
\\&-\frac{c_2 c_3^2  c_1'\ k}{\sqrt{3} c_1 c_2'\ \kt }
-\frac{2\sqrt{3} c_1 c_3^4 K\  k }{D\ \kt}-\frac{k\ \sqrt{3}c_3^2\left((c_2')^2 c_1-c_3^2c_1k^2
-c_1' c_2' c_2\right)}{\kt\ D}
\biggr)\ F_2^{(0)}\\
&+\frac{c_2 c_3^2\  k }{\sqrt{3} c_2'\ \kt}\ \del_r F_2^{(0)}
-\frac{c_2c_3^3k}{D}\ \sum_{\alpha=1}^q
\bigg\{\sqrt{Z_\alpha} a_\alpha'\ F^{(0)}_{(1,\alpha)}\bigg\}
\,,
\end{split}
\eqlabel{h02}
\end{equation}
\begin{equation}
\begin{split}
\h2_{xr}&=\frac{c_2^2c_3^2c_1 }{D}\ \sum_{j=1}^p \bigg\{ \sqrt{\frac{\eta_j}{2}}\phi_j'\  F_{(0,j)}^{(0)}\bigg\}
+\biggl(
-\frac{3\sqrt{3} c_1 c_2 c_2' c_3^2  K}{k \kt\ D}
+ \frac{c_2 c_3^2\ k   }{2\sqrt{3}c_2'D\ \kt }\biggl(c_3^2  c_1k^2\\&
+3 (c_2')^2 c_1+3 c_1' c_2' c_2\biggr)\biggr)\ F_2^{(0)}
+\frac{\sqrt{3} c_2^2}{2k \kt}\  \del_rF_2^{(0)}
-\frac{(c_2^3)'c_3}{2kD}\ \sum_{\alpha=1}^q
\bigg\{\sqrt{Z_\alpha} a_\alpha'\ F^{(0)}_{(1,\alpha)}\bigg\}
\,,
\end{split}
\eqlabel{h03}
\end{equation}
\begin{equation}
\begin{split}
\at^\alpha&=a'_\alpha\ \biggl(
\frac{c_2}{\sqrt{12}k\kt c_2' D}\biggl(c_1c_3^2 k^4+3c_1(c_2')^2(k^2-6K)
+3 c_1'c_2' c_2k^2\biggr)\ F_2^{(0)}\\
&-\frac{(c_2^3)'}{2c_3k D}\ \sum_{\beta=1}^q
\bigg\{\sqrt{Z_\beta} a_\beta'\ F^{(0)}_{(1,\beta)}\bigg\}
+\frac{c_1c_2^2}{2D}\ \sum_{j=1}^p\bigg\{\sqrt{2\eta_j}\phi_j'\  F_{(0,j)}^{(0)}\bigg\}
\biggr)-\frac{c_1c_2}{k c_3\sqrt Z_\alpha}\ \del_r F_{(1,\alpha)}^{(0)}\\
&-\frac{1}{\sqrt{Z_\alpha}}\biggl(\frac{2c_1c_2'}{kc_3}+
\frac{c_1c_2}{2k c_3}\ [\ln Z_\alpha]'
\biggr)\ F_{(1,\alpha)}^{(0)}
\end{split}
\eqlabel{at}
\end{equation}
\begin{equation}
\begin{split}
\ar^\alpha&=-\frac{\sqrt 3 c_2^2 a_\alpha'}{2k\kt c_1^2}\ \del_t  F_2^{(0)}
-\frac{c_2c_3}{kc_1\sqrt{Z_\alpha}}\ \del_t F^{(0)}_{(1,\alpha)}\,,
\end{split}
\eqlabel{ar}
\end{equation}
where we set
\begin{equation}
\kt\equiv \sqrt{k^2-3K}\,,\qquad D\equiv
c_3^2  c_1k^2-3 (c_2')^2 c_1+3 c_1' c_2' c_2\,.
\eqlabel{h0set}
\end{equation}
Additionally,
\begin{equation}
\begin{split}
&\h2_{tt}=\frac{2  c_1^2}{c_3^2} \left(\ln\frac{c_3}{c_1c_2}\right)'\ \h2_{xr}
-\frac{2 c_1^2}{c_3^2}\ \del_r\h2_{xr}+\frac{c_1^2}{c_3^2}\ \h2_{rr}\,,\\
&\h2_{tr}=\frac{4 c_2^2}{k^2}\ \sum_{j=1}^p\bigg\{
\eta_j\phi_j'\ \del_t \f0_j \bigg\}
+2 \del_t\h2_{xr}-\frac{6 c_2' c_2 }{c_3^2\ k^2}\ \del_t\h2_{rr}\,.
\end{split}
\eqlabel{h04}
\end{equation}

The master scalars $\Phi^{(0)}_s$ satisfy \eqref{eqms} with the potentials:
\begin{equation}
\begin{split}
&W_{2,2}^{(0)}=\frac{4}{3D^2 c_2^2} \biggl(
c_2^2 c_1^2 c_3^2\ k^2 \tk^2\ \sum_{n=1}^p \bigg\{\eta_n (\phi_n')^2\bigg\}
+(k^2-6 K) D^2-9\tk^2\  c_1' c_2'c_2  D \\
&-3\tk^2(k^2-2 K)\ c_1 c_3^2 D+2\tk^4\ c_1 c_3^2
\left(c_3^2 k^2 c_1+3 c_1' c_2' c_2\right)+\frac 92 \kt^2 c_2^2(c_2')^2\
\sum_{\gamma=1}^q Z_\gamma (a_\gamma')^2  
\biggr)\,,
\end{split}
\eqlabel{w22}
\end{equation}
\begin{equation}
\begin{split}
&W_{2,(1,\alpha)}^{(0)}=-\frac{\sqrt 3 \kt c_2'}{D c_3}\ \sqrt{Z_\alpha} a'_\alpha\ \biggl( [\ln Z_\alpha]'+\frac{3c_2c_2'}{c_1 D}
\sum_{\gamma=1}^q \biggl\{Z_\gamma (a_\gamma')^2\biggr\}\\
&+\frac{2c_1c_2c_3^2k^2}{3c_2'D} \sum_{n=1}^p\biggl\{\eta_n(\phi_n')^2\biggr\}
+\frac{2}{3c_1c_2c_2'D}\biggl(
c_1^2c_3^4k^4+3c_1^2c_3^2(c_2')^2(k^2-6K)\\&+9c_2(c_2')^2c_1'(c_2'c_1-c_1'c_2)
\biggr)
\biggr)\,,
\end{split}
\eqlabel{w21a}
\end{equation}
\begin{equation}
\begin{split}
&W_{2,(0,i)}^{(0)}=-\frac{\tk  k\  c_3^2 c_1 }{D}\
\sqrt{\frac{2}{3\eta_i}}\ \del_iV
-\frac{\tk k\ c_2 c_1  (c_1' c_2-c_2' c_1)}{D^2}\  \sqrt{\frac{8\eta_i}{3}}\ \phi_i'\ 
\sum_{n=1}^p \bigg\{\eta_n (\phi_n')^2\bigg\}
\\&
+\frac{\tk k\  c_1 c_2'  (D-c_3^2 k^2 c_1) }{c_2 D^2}\  \sqrt{\frac{8\eta_i}{3}}\ \phi_i'
+\frac{\tk k\ c_1^2     c_3^2 c_2'(k^2-2 K)}{D^2c_2}\  \sqrt{24\eta_i}\ \phi_i'
\\
&+\frac{k \kt c_2c_2'}{D^2}\ \sqrt {6\eta_i}\phi_i'\
\sum_{\gamma=1}^q \biggl\{Z_\gamma(a'_\gamma)^2\biggr\}+\frac{k\kt}{\sqrt{6\eta_i}c_1D}
\sum_{\gamma=1}^q\biggl\{(a_\gamma')^2\del_i Z_\gamma \biggr\}
\,,
\end{split}
\eqlabel{w201}
\end{equation}
for $\alpha\ne \beta$:
\begin{equation}
\begin{split}
&W_{(1,\alpha),(1,\beta)}^{(0)}=\frac{a_\alpha'\sqrt{Z_\alpha}\ a_\beta'\sqrt{Z_\beta}}{c_1^2c_3^2D^2} 
\biggl(
c_1^2c_3^4k^4+9c_1^2c_3^2(c_2')^2(k^2-2K)+9(c_2')^2(2c_2'c_1-c_1'c_2)\\&
\times (c_1'c_2-c_2'c_1)+c_1^2c_2^2c_3^2k^2\sum_{n=1}^p\biggl\{\eta_n(\phi_n')^2\biggr\}
+\frac 92 c_2^2 (c_2')^2\sum_{\gamma=1}^q\bigg\{Z_\gamma (a_\gamma')^2\bigg\}\\
&+\frac 32 c_1c_2c_2'D\ \biggl([\ln Z_\alpha]'+[\ln Z_\beta]'\biggr)
\biggr)\,,
\end{split}
\eqlabel{w1a1b}
\end{equation}
and: 
\begin{equation}
\begin{split}
&W_{(1,\alpha),(1,\alpha)}^{(0)}=
\frac{(a'_\alpha)^2Z_\alpha}{c_1^2c_3^2 D^2}
\biggl(
c_1^2c_3^4k^4+9c_1^2c_3^2(c_2')^2(k^2-2K)+9(c_2')^2(2c_2'c_1-c_1'c_2)\\&
\times (c_1'c_2-c_2'c_1)+c_1^2c_2^2c_3^2k^2\sum_{n=1}^p\biggl\{\eta_n(\phi_n')^2\biggr\}
+\frac 92 c_2^2 (c_2')^2\sum_{\gamma=1}^q\bigg\{Z_\gamma (a_\gamma')^2\bigg\}\\
&+3 c_1c_2c_2'D\ [\ln Z_\alpha]'
\biggr)+\frac{2}{3c_3^2}\sum_{n=1}^p\bigg\{\eta_n (\phi_n')^2
\bigg\}
-\frac{4c_1'c_2'}{c_1c_2c_3^2}
-\frac{[\ln Z_\alpha]''}{2c_3^2}+\frac{([\ln Z_\alpha]')^2}{4c_3^2}\\
&+\frac{c_1c_2c_3'-c_3(c_1'c_2-c_2'c_1)}{2c_1c_2c_3^3}\ [\ln Z_\alpha]'\,,
\end{split}
\eqlabel{w1a1a}
\end{equation}
\begin{equation}
\begin{split}
&W_{(1,\alpha),(0,i)}^{(0)}=\frac{ka_\alpha'\sqrt{Z_\alpha}\sqrt{2\eta_i}}{D} 
\biggl(\frac{c_2c_3}{2\eta_i}\ \del_iV+\frac{\phi_i'c_2^2(c_1'c_2-c_2'c_1)}{c_3D}
\sum_{n=1}^p\bigg\{\eta_n(\phi'_n)^2\bigg\}\\
&-\frac{3\phi_i'c_2^2c_2'}{2c_1c_3D}\sum_{\gamma}\bigg\{Z_\gamma(a'_\gamma)^2\biggr\}
+\frac{\phi_i'c_2'}{c_3D}\biggl(c_1c_3^2(6K-5k^2)+9c_2'(c_2'c_1-c_1'c_2)\biggr)\\
&-\frac{\phi'_ic_2(k^2c_1c_3^2+3c_2'(c_2c_1'-c_1c_2'))}{2c_3D}\
[\ln Z_\alpha]'
-\frac{c_2}{4\eta_ic_1^2c_3}\sum_{\gamma=1}^q
\bigg\{(a_\gamma')^2\del_i Z_\gamma\bigg\}\\
&-\frac{D}{2\eta_ic_1c_2c_3}\ \del_i\ln Z_{\alpha}
\biggr)\,,
\end{split}
\eqlabel{w1a0i}
\end{equation}
finally, for any $i$ and $j$:
\begin{equation}
\begin{split}
&W_{(0,i),(0,j)}^{(0)}=\frac{1}{2 \sqrt{\eta_i} \sqrt{\eta_j}}\ \del_i\del_jV
-\frac{c_2 (c_2' c_1-c_1' c_2)}{D} \left(
\sqrt{\frac{\eta_j}{\eta_i}}\ \phi_j'\ \del_iV
+\sqrt{\frac{\eta_i}{\eta_j}}\ \phi_i'\ \del_jV\right)
\\&+\frac{2 c_2^2 (c_2' c_1-c_1' c_2)^2}{c_3^2 D^2}\
\sqrt{\eta_i}\phi_i'\ \sqrt{\eta_j}\phi_j'\  \sum_{n=1}^p \bigg\{\eta_n (\phi_n')^2\bigg\}
+\frac{\sqrt{\eta_i}\phi_i'\ \sqrt{\eta_j}\phi_j'}{D^2}
\biggl(2 c_1^2 c_3^2\ k^2 (k^2-2 K)\\
&+2k^2\ (c_2' c_1+2 c_1' c_2) (c_2' c_1-c_1' c_2)
-\frac{12 c_2' (c_2' c_1+c_1' c_2)
(c_2' c_1-c_1' c_2)^2}{c_3^2 c_1}\biggr)
\\&-\frac{1}{4c_1^2c_3^2\sqrt \eta_i\sqrt \eta_j}\sum_{\gamma=1}^q\bigg\{
(a'_\gamma)^2 \del^2_{ij}Z_{\gamma}\biggr\}
+\frac{c_2(c_1c_2'-c_2c_1')}{2c_1^2c_3^2D}
\biggl(
\sqrt\frac{\eta_i}{\eta_j}\ \phi_i'\ \sum_{\gamma=1}^q \bigg\{
(a'_\gamma)^2 \del_j Z_\gamma\bigg\}\\
&+\sqrt\frac{\eta_j}{\eta_i}\ \phi_j'\ \sum_{\gamma=1}^q \bigg\{
(a'_\gamma)^2 \del_i Z_\gamma\bigg\}
\biggr)+\frac{1}{2c_1^2c_3^2\sqrt\eta_i\sqrt\eta_j}\sum_{\gamma=1}^q
\bigg\{(a'_\gamma)^2\ \del_i\ln Z_\gamma\ \cdot \del_jZ_\gamma\bigg\}
\,,
\end{split}
\eqlabel{w0102}
\end{equation}
where $\tk$ and $D$ are defined in \eqref{h0set}.

Notice that for $K>0$, the relations  \eqref{h02} and \eqref{h03}  between the master scalars
and the gauge invariant fluctuations are singular for $\ell=0$ and $\ell=1$.  
This is because in these cases the dynamical degrees of freedom are those of the
scalars only (for $\ell=0$)
and scalars and vectors (for $\ell=1$).
These cases must be treated separately \cite{Jansen:2019wag}.

\subsubsection{$\ell=0$}\label{h0l0s}

For $\ell=0$, there are no $h_{tx}$, $h_{xr}$ and $h_-$ components of the metric perturbations,
and no $a_x^\alpha$ components of the vector perturbations;
furthermore, the gauge transformations can be used to set metric components $h_{tr}=0$ and $h_+=0$, and the vector components $a_t^\alpha=0$.
Thus, we are left with perturbations
\begin{equation}
\{\f0_j\,,\ a_r^\alpha\,,\ h_{tt}\,,\ h_{rr}\}\,.
\eqlabel{h0l0}
\end{equation}
We find\footnote{As pointed out in \cite{Jansen:2019wag}, here, as well as for $\ell=1$,
there is also a certain inhomogeneous
piece in the master equations. This piece must be set to zero to study  fluctuations
in a fixed-mass black hole background.}
\begin{equation}
\begin{split}
&h_{rr}=\frac{2c_3^2  c_2}{3c_2'}\ \sum_{i=1}^p \bigg\{\eta_i \phi_i' \f0_i\bigg\}\,,
\end{split}
\eqlabel{l01}
\end{equation}
\begin{equation}
\begin{split}
&c_1^2\ \del_r\left(\frac{h_{tt}}{c_1^2}\right)=
-\frac{2 c_1^2 c_2 }{3c_2'}\   \sum_{i=1}^p\bigg\{\eta_i\phi_i'\ \del_r\f0_i\bigg\}
+\frac{c_1^2 c_3^2 c_2 }{3c_2'} \sum_{i=1}^p\bigg\{\del_iV\ \f0_i\bigg\}
\\&+\frac{2 c_1^2 c_2^2} {9(c_2')^2}\
\sum_{j=1}^p\bigg\{\eta_j(\phi_j')^2\bigg\}\ \sum_{i=1}^p\bigg\{\eta_i \phi_i'\  \f0_i\bigg\}
-\frac{4 c_1 (c_1 c_2'+c_2 c_1')}{3c_2'}
\ \sum_{i=1}^p\bigg\{ \eta_i\phi_i' \ \f0_i\bigg\}\\
&-\frac{c_2}{6c_2'}\sum_{i=1}^p\sum_{\alpha=1}^q\bigg\{(a_\alpha')^2\del_i Z_\alpha\ \f0_i\biggr\}\,,
\end{split}
\eqlabel{l02}
\end{equation}
\begin{equation}
\begin{split}
&\del_t a_r^\alpha=a_\alpha'\biggl(
\frac{h_{tt}}{2c_1^2}-\frac{c_2}{3c_2'}\sum_{i=1}^p \bigg\{\eta_i \phi_i' \f0_i\bigg\}+
\sum_{i=1}^p \bigg\{\del_i \ln Z_\alpha\ \f0_i\bigg\}
\biggr)\,.
\end{split}
\eqlabel{l0h0ar}
\end{equation}
Introducing the master scalars as 
\begin{equation}
\f0_j=-\frac{1}{\sqrt{2\eta_j}}\ F^{(0)}_{(0,j)}\,,
\eqlabel{h0l0def}
\end{equation}
we obtain master equations for $\Phi^{(0)}_{(0,j)}$ with $W^{(0)}_{(0,i),(0,j)}$
given formally by \eqref{w0102} in the limit $k\to 0$.

\subsubsection{$\ell=1$}

For $\ell=1$ there is no $h_-$ component of the metric fluctuations. 
We can use gauge transformations to set metric components $h_{tx}=0$ and
$h_+=0$. Thus, are left with the (gauge variant) perturbations 
\begin{equation}
\{\f0_j\,,\ \at^\alpha\,,\ \ar^\alpha\,,\ h_{tr}\,,\ h_{rr}\,,\ h_{xr}\,,\ h_{tt}\}\,.
\eqlabel{gauge01}
\end{equation}
Note that the scalar eigenfunction for $\ell=1$ is explicitly
\begin{equation}
S(X_{3,K})=x \sqrt{K}\,.
\eqlabel{sl1}
\end{equation}
From the fluctuation equations of motion we find (compare with \eqref{h04})
\begin{equation}
\begin{split}
&h_{tt}=\frac{2  c_1^2}{c_3^2} \left(\ln\frac{c_3}{c_1c_2}\right)'\ h_{xr}
-\frac{2 c_1^2}{c_3^2}\ \del_rh_{xr}+\frac{c_1^2}{c_3^2}\ h_{rr}\,,\\
&h_{tr}=\frac{4 c_2^2}{k^2}\ \sum_{j=1}^p\bigg\{
\eta_j\phi_j'\ \del_t \f0_j \bigg\}
+2 \del_t h_{xr}-\frac{6 c_2' c_2 }{c_3^2\ k^2}\ \del_t h_{rr}\,.
\end{split}
\eqlabel{htr}
\end{equation}

The remaining fluctuations can be expressed through the master scalars \eqref{gauge0m} as
\begin{equation}
\begin{split}
\f0_j&=-\frac{1}{\sqrt{2\eta_j}}\ F_{(0,j)}^{(0)}+\frac{g\ c_2\phi_j'}
{c_2'} \ {F}_2^{(0)} \,,
\end{split}
\eqlabel{h01l1}
\end{equation}
\begin{equation}
\begin{split}
&h_{rr}= \frac{ c_2 c_3^2  (c_2' c_1-c_1' c_2)}{D}\ \sum_{j=1}^p\bigg\{\sqrt{2\eta_i}\phi_j'\
F_{(0,j)}^{(0)}\bigg\}
+2g \biggl(
\frac{c_2^2 c_3^2}{3(c_2')^2}\ \sum_{j=1}^p \bigg\{\eta_j (\phi_j')^2\bigg\}
\\&-\frac{c_2 c_3^2  c_1'}{c_1 c_2'}
-\frac{6 c_1 c_3^4 K }{D}-\frac{3c_3^2\left((c_2')^2 c_1-c_3^2c_1k^2
-c_1' c_2' c_2\right)}{D}
\biggr)\ F_2^{(0)}+2g\ \frac{c_2 c_3^2 }{ c_2'}\ \del_r F_2^{(0)}\\
&-\frac{c_2c_3^3k \sqrt 2}{D}\sum_{\alpha=1}^q\bigg\{a'_\alpha\sqrt{Z_\alpha}\ F^{(0)}_{(1,\alpha)}\bigg\}\,,
\end{split}
\eqlabel{h02l1}
\end{equation}
\begin{equation}
\begin{split}
h_{xr}&=\frac{c_2^2c_3^2c_1 }{D}\ \sum_{j=1}^p \bigg\{ \sqrt{\frac{\eta_j}{2}}\phi_j'\  F_{(0,j)}^{(0)}\bigg\}
+2g \biggl(
-\frac{9 c_1 c_2 c_2' c_3^2  K}{k^2\ D}
+ \frac{c_2 c_3^2   }{2c_2'D }\biggl(c_3^2  c_1k^2\\&
+3 (c_2')^2 c_1+3 c_1' c_2' c_2\biggr)\biggr)\ F_2^{(0)}
+g\ \frac{{3} c_2^2}{k^2}\  \del_rF_2^{(0)}-\frac{(c_2^3)'c_3}{\sqrt 2 k D}
\sum_{\alpha=1}^q\bigg\{a'_\alpha\sqrt{Z_\alpha}\ F^{(0)}_{(1,\alpha)}\bigg\}\,,
\end{split}
\eqlabel{h03l1}
\end{equation}
\begin{equation}
\begin{split}
\at^\alpha&=a'_\alpha\ \biggl(
\frac{c_2}{c_2'}\ g\ F_2^{(0)}
-\frac{(c_2^3)'}{2c_3k D}\ \sum_{\beta=1}^q
\bigg\{\sqrt{Z_\beta} a_\beta'\ F^{(0)}_{(1,\beta)}\bigg\}
+\frac{c_1c_2^2}{2D}\ \sum_{j=1}^p\bigg\{\sqrt{2\eta_j}\phi_j'\  F_{(0,j)}^{(0)}\bigg\}
\biggr)\\
&-\frac{c_1c_2}{k c_3\sqrt Z_\alpha}\ \del_r F_{(1,\alpha)}^{(0)}
-\frac{1}{\sqrt{Z_\alpha}}\biggl(\frac{2c_1c_2'}{kc_3}+
\frac{c_1c_2}{2k c_3}\ [\ln Z_\alpha]'
\biggr)\ F_{(1,\alpha)}^{(0)}\,,
\end{split}
\eqlabel{atl1}
\end{equation}
\begin{equation}
\begin{split}
\ar^\alpha&=-\frac{3 c_2^2 a_\alpha'}{k^2 c_1^2}\ g\ \del_t  F_2^{(0)}
-\frac{c_2c_3}{kc_1\sqrt{Z_\alpha}}\ \del_t F^{(0)}_{(1,\alpha)}\,,
\end{split}
\eqlabel{arl1}
\end{equation}
where $D$ is given by \eqref{h0set}, and $g$ is an arbitrary constant parameter of the unfixed
gauge transformation  \cite{Jansen:2019wag}.
Note that \eqref{h01l1}-\eqref{arl1} are equivalent to  \eqref{h01}-\eqref{ar} up to
replacement
\begin{equation}
2g\ \longleftrightarrow\ \frac{k}{\sqrt3\ \tk}\,.
\eqlabel{replacementh0}
\end{equation}

The only physical master equations are those for the scalars $\Phi^{(0)}_{(0,j)}$
and  $\Phi^{(0)}_{(1,\alpha)}$
(in these equations there is a decoupling of the 'gauge' master scalar $\Phi^{(0)}_2$), with the
relevant potentials $W^{(0)}_{(1,\alpha),(1,\beta)}$, $W^{(0)}_{(1,\alpha),(0,j)}$ and $W^{(0)}_{(0,i),(0,j)}$
obtained setting $\tk=0$ in \eqref{w1a1b}, \eqref{w1a0i} and \eqref{w0102}.
The equation for the 'gauge' master scalar can also be obtained from the general expressions
valid for $\ell\ge 2$, provided we identify (compare \eqref{h01} and  \eqref{h01l1})
\begin{equation}
F_2^{(0)}\bigg|_{\ell\ge 2}\equiv 2g\ \frac{\sqrt3\ \tk}{k}\ F_2^{(0)}\bigg|_{\ell=1}\,,
\eqlabel{replacementf2}
\end{equation}
prior to taking the limit $\tk \to 0$.
Because of \eqref{replacementf2}, this latter equation is necessarily singular in the limit $g\to 0$.

\subsection{Helicity $h=1$ sector}

The relation between the gauge invariant fluctuations in the vector sector
\begin{equation}
\{\az^\alpha\,,\ \h2_{tz}\,,\ \h2_{zr}\}\,,
\eqlabel{gauge1}
\end{equation}
and the master scalar
\begin{equation}
\{F^{(1)}_{(1,\alpha)}\,,\ F^{(1)}_2\}\,,
\eqlabel{gauge1m}
\end{equation}
all being functions of $(t,r)$, is as follows:
\begin{equation}
\begin{split}
&\h2_{tz}=\frac{3 c_2 c_1 c_2'}{\tk\ c_3}\ F^{(1)}_2+\frac{c_2^2 c_1}{\tk\ c_3}\ \del_r F^{(1)}_2\,,
\\
&\h2_{zr}=\frac{c_3 c_2^2}{\tk\ c_1}\ \del_tF^{(1)}_2 \,,\\
&\az^\alpha=\frac{c_2}{\sqrt Z_\alpha}\ F^{(1)}_{(1,\alpha)}\,,
\end{split}
\eqlabel{h12}
\end{equation}
where we deliberately introduced a singularity at $\ell=1$\, \ie $\kt=0$ \eqref{h0set},  
to highlight the fact that the fluctuations \eqref{gauge1} are gauge invariant only for $\ell\ge 1$
\cite{Jansen:2019wag}. 

The master scalar $\Phi^{(1)}_2$ satisfies \eqref{eqms} with the potentials
\begin{equation}
W^{(1)}_{2,2}=\frac{1}{c_3^2}\ \sum_{j=1}^p \bigg\{\eta_j(\phi_j')^2\bigg\}-
\frac{3 K}{c_2^2}+\frac{3 (c_2')^2}{c_2^2 c_3^2}-\frac{6 c_1'c_2'}
{c_1 c_2 c_3^2}\,,
\eqlabel{w122}
\end{equation}
\begin{equation}
\begin{split}
&W^{(1)}_{2,(1,\alpha)}=-\frac{\kt a_\alpha'\sqrt Z_\alpha}{c_1c_2c_3}\,,
\end{split}
\eqlabel{w12a}
\end{equation}
for $\a\ne \beta$,
\begin{equation}
\begin{split}
&W^{(1)}_{(1,\alpha),(1,\beta)}=\frac{a_\alpha'\sqrt Z_\alpha\
a_\beta'\sqrt Z_\beta}{c_1^2c_3^2}\,,
\end{split}
\eqlabel{w11a1b}
\end{equation}
and,
\begin{equation}
\begin{split}
&W^{(1)}_{(1,\alpha),(1,\alpha)}=\frac{(a'_\alpha)^2Z_\alpha}{c_1^2c_3^2}
+\frac{[\ln Z_\alpha]''}{2c_3^2}
+\frac{([\ln Z_\alpha]')^2}{4 c_3^2}+\frac{[\ln Z_\alpha]'}{2c_3^2}\left(
\ln\frac{c_1c_2}{c_3}\right)'\\
&+\frac{1}{3c_3^2} \sum_{n=1}^p\bigg\{\eta_n(\phi_n')^2\bigg\}
+\frac{K c_1c_3^2-c_1(c_2')^2-2c_2c_1'c_2'}{c_1c_2^2c_3^2}\,.
\end{split}
\eqlabel{w11a1a}
\end{equation}

\subsubsection{$\ell=1$}

The case $\ell=1$ is special since in this case there are no dynamical degrees of freedom in the metric, only in the gauge fields
\cite{Jansen:2019wag}. 
Indeed, here, there are no $h_{xz}$ component of the metric fluctuations,  and the gauge can be fixed
setting $h_{zr}=0$.
The remaining metric component $h_{tz}$ satisfies
\begin{equation}
c_2\ \del_r \left(\frac{h_{tz}}{c_2^2}\right)=-\sum_{\alpha=1}^q\bigg\{
a_\alpha'\sqrt Z_\alpha\ F^{(1)}_{(1,\alpha)}
\bigg\}\,,
\eqlabel{htz}
\end{equation}
where the master scalars $F^{(1)}_{(1,\alpha)}$
for the gauge fields are introduced as in \eqref{h12}. The master equations
for $\Phi^{(1)}_{(1,\alpha)}$ are with
$W^{(1)}_{(1,\alpha),(1,\beta)}$ given by \eqref{w11a1b} and
\eqref{w11a1a}, with $\kt\to 0$. Notice that in this limit
all $W^{(1)}_{2,(1,\alpha)}$ vanish.

\subsection{Helicity $h=2$ sector}

The relation between a particular  gauge invariant fluctuation $\h2_{yz}$ and the
master  scalar $F_2^{(2)}$ in the tensor sector is as follows 
\begin{equation}
\delta ds_5^2=\delta g_{yz}(t,r,X_{3,K})\ dydz ={\h2_{yz}(t,r)}\ \frac{S_T(x)}{1-K y^2}\ dydz=
c_2^2 F_2^{(2)}(t,r)\ \frac{S_T(x)}{1-K y^2}\ dydz  \,,
\end{equation}
where $S_T(x)$ satisfies \cite{Jansen:2019wag}
\begin{equation}
(1-K x^2)\ S_T''=-K x\ S_T'-\left(k^2+2 K \left(1+\frac{K x^2}{1-K x^2}\right)\right)\ S_T\,.
\eqlabel{sneq}
\end{equation}

The master scalar
\begin{equation}
\Phi_2^{(2)}(\xi)\equiv F_2^{(2)}(t,r)\ S(X_{3,K}) 
\eqlabel{msh2}
\end{equation}
equation is {\it universally} that of the minimally coupled massless scalar
\begin{equation}
\square \Phi_2^{(2)}=0\,,
\eqlabel{eqms2}
\end{equation}
\ie
\begin{equation}
W^{(2)}_{2,2}\equiv 0\,.
\eqlabel{w222}
\end{equation}
The universality of the $h=2$ sector at $K=0$ was emphasized originally in \cite{Buchel:2004qq,Benincasa:2006fu}.

\section{The STU model}\label{stuqnm}

The STU model is a consistent truncation of type IIB supergravity
on five-sphere \cite{Behrndt:1998jd}: 
\begin{equation}
\begin{split}
S_{eff}=\frac{1}{2\kappa_5^2} \int_{\calm_5} \biggl(
&R-\frac 14 G_{ab}F_{\rho\sigma}^a F_{\mu\nu}^b g^{\rho\mu}g^{\sigma\nu}+
\frac{c_{abc}}{48 \sqrt{2}}\epsilon^{\mu\nu\rho\sigma\lambda}F^a_{\mu\nu}F_{\rho\sigma}^b A^c_\lambda
\\&
\qquad -G_{ab} g^{\mu\nu}\del_\mu X^a\del_\nu X^b+\sum_{a=1}^3 \frac{4}{X^a}
\biggr) \star 1\,,
\end{split}
\eqlabel{seff}
\end{equation}
where $c_{abc}$ are symmetric constants, nonzero only for distinct indexes with $c_{123}=1$, $g_{\mu\nu}$ is the metric on $\calm_5$, $F^a_{\mu\nu}$ are the field strengths for the gauge fields $A^a_\mu$, $a=1\cdots 3$,
dual to conserved currents of the maximal Abelian subgroup of the $SU(4)$ $R$-symmetry of $\caln=4$ SYM. The three
real positive neutral scalar fields $X^a$  describe the deformation of $S^5$ in the uplift of $S_{eff}$ to type IIB supergravity; they are
constrained, at the level of the effective action \eqref{seff}, by
\begin{equation}
X^1 X^2 X^3=1\,.
\eqlabel{constr}
\end{equation}
The field space metric $G_{ab}$ is
\begin{equation}
G_{ab}=\frac 12 {\rm diag} \biggl((X^1)^{-2}\,,\, (X^2)^{-2}\,,\, (X^3)^{-2}\biggr)\,.
\eqlabel{const}
\end{equation}
The gravitational constant $\kappa_5$ is related to the central charge $c$ of the boundary gauge theory
as
\begin{equation}
\kappa_5^2=\frac{\pi^2}{c}=\frac{4\pi^2 }{N_c^2}\,.
\eqlabel{k5}
\end{equation}
Effective action \eqref{seff} allows for analytic solutions of black holes in asymptotically $AdS_5$ with
distinct $U(1)^3$ charges \cite{Behrndt:1998jd}, realizing the gravitational dual to charged $\caln=4$ SYM plasma.

We proceed to review the background geometry and the thermodynamics of the
STU black holes in section \ref{stubt}, and discuss the helicity $h=0$ fluctuations
about their symmetric phase in section \ref{stufluc}. 

\subsection{The STU black hole background geometry}\label{stubt}

The effective action \eqref{seff} can put put in the form \eqref{efac} with
the identification \cite{Cvetic:1999xp},
\begin{equation}
\begin{split}
X^\alpha\equiv e^{-\frac 12 \vec{b}_\alpha \vec \phi}\,,\qquad \vec\phi=\begin{bmatrix}
\phi_1\\\phi_2
\end{bmatrix}\,,\ \vec{b}_1=\begin{bmatrix}
\frac{2}{\sqrt 6}\\\sqrt 2
\end{bmatrix}
\,,\ \vec{b}_2=\begin{bmatrix}
\frac{2}{\sqrt 6}\\-\sqrt 2
\end{bmatrix}
\,,\ \vec{b}_3=\begin{bmatrix}
-\frac{4}{\sqrt 6}\\0
\end{bmatrix}\,,
\end{split}
\eqlabel{stugen}
\end{equation}
where the constraint \eqref{constr} is insured by $\sum_\alpha\vec{b}_\alpha=\vec 0$,  and 
\begin{equation}
\eta_i=\frac 12\,,\qquad V(\{\phi_i\})=-4\sum_{\alpha=1}^3  e^{\frac 12 \vec{b}_\alpha \vec \phi}\,,\qquad
Z_\alpha(\{\phi_i\})=\frac 12  e^{\vec{b}_\alpha \vec \phi}\,.
\eqlabel{stugen1}
\end{equation}
The background equations of motion \eqref{bac1}-\eqref{bac4} can be solved analytically
\cite{Behrndt:1998jd},
\begin{equation}
\begin{split}
&X^\alpha=\frac{H^{1/3}}{H^\alpha}\,,\quad H=\prod_{a=1}^3 H^\alpha\,,\quad H^\alpha=1
+\kappa_\alpha r\,,
\quad a_\alpha=\frac{\sqrt{2\kappa_\alpha}\ \sqrt{\beta+K\kappa_\alpha}}
{(1+\kappa_\alpha)(1+\kappa_\alpha r)}(1-r)\,,
\\
&c_1=\frac{(\pi T_0) \sqrt f}{H^{1/3}\sqrt r}\quad
c_2=\frac{(\pi T_0) H^{1/6}}{\sqrt r}\,,\quad c_3=\frac{H^{1/6}}{2r\sqrt f}\,,
\quad f=H+\frac{K r}{(\pi T_0)^2}-\frac{\beta r^2}{(\pi T_0)^2}\,,
\end{split}
\eqlabel{stubac}
\end{equation}
where the constant $\beta$ is chosen so that the black hole horizon is
at $r=1$,
\begin{equation}
\beta=K+(\pi T_0^2)\ \prod_{a=1}^3(1+\kappa_\alpha)\,.
\eqlabel{defb}
\end{equation}
The parameters of the solution, $T_0$ and $\kappa_\alpha$, determine
the $U(1)$ chemical potentials $\mu_\alpha$ and the
corresponding charge densities $\rho_\alpha$, the temperature $T$,
the  entropy $s$, the energy density $\cale$, and the Gibbs free energy density $\Omega$
of the STU black hole as follows  ( see \eg \cite{Harmark:1999xt,Cvetic:1999ne}),
\begin{equation}
\begin{split}
&\mu_\alpha=\frac{\sqrt{2\kappa_\alpha}\ \sqrt{\beta+K\kappa_\alpha}}
{1+\kappa_\alpha}\,,\qquad  \rho_\alpha=\frac{N_c^2 T_0^2}{8}\ \sqrt{2\kappa_\alpha}\ \sqrt{\beta+K\kappa_\alpha}\,,\\
&T=\frac{K+(\pi T_0)^2\ (2+\sum_\alpha\kappa_\alpha
-\prod_\alpha\kappa_\alpha)}{2\pi^2T_0\ \prod_\alpha (1+\kappa_\alpha)^{1/2}}\,,\qquad
s=\frac{\pi^2 N_c^2T_0^3}{2}\ \prod_\alpha(1+\kappa_\alpha)^{1/2}\,,\\
&\cale=\frac{N_c^2T_0^2}{4}\left(K\sum_\alpha\kappa_\alpha+\frac 32\beta\right)\,,\qquad
\Omega=\cale-s T -\sum_\alpha \mu_\alpha\cdot \rho_\alpha\,.
\end{split}
\eqlabel{stuthermo}
\end{equation}
We can readily verify from \eqref{stuthermo} the first law of thermodynamics
\begin{equation}
dE=T\cdot dS +\sum_\alpha \mu_\alpha\cdot dQ_\alpha-P\cdot dV\,,
\eqlabel{1stlaw}
\end{equation}
where
\begin{equation}
V={\rm vol}(S^3)=\frac{2\pi^2}{K^{3/2}}\,,\qquad E=\cale V\,,\qquad Q_\alpha=\rho_\alpha V\,,\qquad  S=s V
\eqlabel{1stlaw1}
\end{equation}
Note that while the conformal symmetry of the $\caln=4$ SYM relates
the pressure $P=\frac \cale3$ of the thermal states of the theory, $\Omega\ne -P$
whenever $K\ne 0$.

\subsection{Thermodynamic stability in the microcanonical ensemble}
\label{stuthermos}
Following \cite{Gladden:2024ssb}, to study the thermodynamic
stability of the STU black holes (correspondingly the thermal states of
$\caln=4$ SYM plasma on $S^3$) we need to obtain the equilibrium equation of
state (EoS) $\cale=\cale(s,\rho_\alpha)$. It is difficult to obtain this EoS analytically at
finite $K$; thus we work perturbatively in $\frac{K}{T^2}$:
\begin{equation}
\cale=\frac{3}{2(2\pi N_c)^{2/3}}\cdot s^{4/3}\cdot \prod_\alpha\left(1
+\frac{8\pi^2\rho_\alpha^2}{s^2}\right)^{1/3}\cdot \sum_{k=0}^\infty \bigg\{ \epsilon_k\cdot
\left(\frac{N_c^2}{2\pi s}\right)^{\frac{2k}{3}}
\cdot K^k\bigg\}\,,
\eqlabel{eosstu}
\end{equation}
with the first couple coefficients $\epsilon_k$ are given explicitly by 
\begin{equation}
\epsilon_1=\frac{1+\sum_\alpha\frac{8\pi^2\rho_\alpha^2}{3s^2}}
{\prod_\alpha\left(1+\frac{8\pi^2\rho_\alpha^2}{s^2}\right)^{2/3}}\,,\qquad 
\epsilon_2=-\frac{64\pi^4}{9}\ \frac{\sum_\alpha\frac{\rho_\alpha^4}{s^4}-\sum\sum_{\alpha> \beta}
\frac{\rho_\alpha^2\rho_\beta^2}{s^4}}
{\prod_\alpha\left(1+\frac{8\pi^2\rho_\alpha^2}{s^2}\right)^{4/3}}\,.
\eqlabel{defepsilon}
\end{equation}
An equilibrium state is thermodynamically stable\footnote{I would like
to thank Pavel Kovtun for discussing the thermodynamic stability.} if the Hessian matrix of the second
derivatives of $E(Q_\alpha,S,V)$, \ie $\calh^E_{ij}\equiv \del^2E/\del y_i\del y_j$,
where $y_i=(Q_\alpha,S,V)$, $\alpha=\{3,2,1\}$, is positive definite \cite{Callen:450289}.
Evaluating the Hessian $\calh^E$ and its leading principle minors
\[
\left\{\Delta_1\equiv \frac{\del^2 E}{\del Q_3^2}\,,\qquad \cdots\,,\qquad 
\Delta_5\equiv \det \calh^E_{ij}\right\}\,,
\] we obtain
the thermodynamic stability conditions in terms $\kappa_\alpha$ and $\frac{K}{(\pi T_0)^2}$.
The full set of these conditions, for $K=0$, is presented in \cite{Gladden:2024ssb}.
These constraints will be modified whenever $K\ne 0$:
\begin{equation}
\begin{split}
\Delta_1:\qquad 0<\biggl(3-\kappa_3\biggr)
+\frac{4}{3\prod_\alpha (1+\kappa_\alpha)}
\biggl(
\kappa_3(\kappa_1+\kappa_2+\kappa_3)
\biggr)\cdot \frac{K}{(\pi T_0)^2}
+\calo\left(\frac{K^2}{(\pi T_0)^4}\right)\,,
\end{split}
\eqlabel{eq4a}
\end{equation}
\begin{equation}
\begin{split}
&\Delta_2:\qquad 0< \biggl(3-\kappa_2-\kappa_3-\kappa_2 \kappa_3\biggr)-\frac{4}{3\prod_\alpha (1+\kappa_\alpha)}
\biggl(
\kappa_2 \kappa_3 (2 \kappa_1-\kappa_2-\kappa_3)\\&-(\kappa_2+\kappa_3)
(\kappa_1+\kappa_2+\kappa_3)\biggr)\cdot \frac{K}{(\pi T_0)^2}
+\calo\left(\frac{K^2}{(\pi T_0)^4}\right)\,,
\end{split}
\eqlabel{eq4b}
\end{equation}
\begin{equation}
\begin{split}
&\Delta_3:\qquad 0< \biggl(
3-\kappa_1-\kappa_2-\kappa_3
-\kappa_1\kappa_2-\kappa_1\kappa_3-\kappa_2\kappa_3
+3\kappa_1\kappa_2\kappa_3
\biggr)\\
&+\frac{4}{3\prod_\alpha (1+\kappa_\alpha)}
\biggl(
(\kappa_1+\kappa_2+\kappa_3)
(\kappa_1+\kappa_2+\kappa_3+\kappa_1\kappa_2+\kappa_1\kappa_3+\kappa_2\kappa_3)
\\&-3\kappa_1\kappa_2\kappa_3(\kappa_1+\kappa_2+\kappa_3+3)
\biggr)\cdot \frac{K}{(\pi T_0)^2}
+\calo\left(\frac{K^2}{(\pi T_0)^4}\right)
\,,
\end{split}
\eqlabel{eq4c}
\end{equation}
\begin{equation}
\begin{split}
&\Delta_4:\qquad 0< \biggl(
2-\kappa_1-\kappa_2-\kappa_3+\kappa_1\kappa_2\kappa_3
\biggr)
-\frac{1}{3\prod_\alpha (1+\kappa_\alpha)}
\biggl(
-(\kappa_1+\kappa_2+\kappa_3)\\
&\times \bigl(4(\kappa_1+\kappa_2+\kappa_3)+1\bigr)
+3(\kappa_1\kappa_2+\kappa_1\kappa_3+\kappa_2\kappa_3)
+\kappa_1\kappa_2\kappa_3\bigl(4(\kappa_1+\kappa_2+\kappa_3)+15\bigr)
\\&+3
\biggr)\cdot \frac{K}{(\pi T_0)^2}
+\calo\left(\frac{K^2}{(\pi T_0)^4}\right)
\,,
\end{split}
\eqlabel{eq4d}
\end{equation}
\begin{equation}
\begin{split}
\Delta_5:\qquad 0< -\biggl(
2-\kappa_1-\kappa_2-\kappa_3+\kappa_1\kappa_2\kappa_3
\biggr)\cdot
\frac{K}{(\pi T_0)^2}
+\calo\left(\frac{K^2}{(\pi T_0)^4}\right)
\,.
\end{split}
\eqlabel{eq4e}
\end{equation}
Notice that \eqref{eq4d} and \eqref{eq4e} are always mutually
incompatible: $U(1)^3$-charged $\caln=4$ SYM plasma is {\it always}
thermodynamically
unstable\footnote{We verified that
when $\kappa_\alpha=0$, \ie  for the neutral plasma, $\caln=4$ SYM 
is always thermodynamically unstable on $S^3$, no matter the size, provide
the $S^3$ volume is allowed to fluctuate.} on $S^3$ (at least for small $K>0$).
The instability statement is contingent about the ability to fluctuate
the volume of the three-sphere: while in the holographic setting the
spatial manifold on which the boundary theory is formulated is fixed,
the microcanonical ensemble where $dV\ne 0$ is a standard ensemble
in statistical mechanics where the corresponding thermodynamic potential is
the total energy $E(Q_\alpha,S,V)$ \cite{LandauLifshitzStatPhys}.
A potential caveat is that it is difficult to imagine a process where
the fluctuating volume remains a three-sphere. If one insists that
the thermodynamic stability is to be analyzed in the microcanonical
ensemble with $dV\equiv 0$, one simply drops the constraint \eqref{eq4e}
above.

The flat space, \ie $K=0$, thermodynamic instability for a
{\it symmetric} state with $\kappa_\alpha\equiv \kappa$ was triggered
from violation of the $\Delta_2$ leading minor condition
\eqref{eq4b}; this condition for the symmetric state becomes  
\begin{equation}
\begin{split}
&0<(\kappa+3) (1-\kappa)+\frac{8 \kappa^2}{(1+\kappa)^3}\cdot \frac{K}{(\pi T_0)^2}
+\frac{4 \kappa^2 (\kappa^2-4 \kappa-2)}{(1+\kappa)^6}\cdot \frac{K^2}{(\pi T_0)^4}\\
&-
\frac{8 \kappa^2 (1+2 \kappa) (\kappa^2-2 \kappa-1)}{(1+\kappa)^9}
\cdot \frac{K^3}{(\pi T_0)^6}
+\frac{4 \kappa^2 (1+2 \kappa)^2 (3 \kappa^2-4 \kappa-2)}{(1+\kappa)^{12}}
\cdot \frac{K^4}{(\pi T_0)^8}\\&+\calo\left(\frac{K^5}{(\pi T_0)^{10}}\right)\,.
\end{split}
\eqlabel{eq4c1}
\end{equation}
Eq.~\eqref{eq4c1} can be rephrased as a minimal curvature $K$ condition, necessary
to enforce the positivity $\Delta_2 >0$ for $\kappa>1$,
\begin{equation}
\begin{split}
\frac{K}{(\pi T_0)^2}\ >\ 4(\kappa-1)+4(\kappa-1)^2+(\kappa-1)^3+ \calo((\kappa-1)^5) \,.
\end{split}
\eqlabel{eq4c2}
\end{equation}

\subsection{Thermodynamic stability in the enthalpy representation ensemble}
\label{stuthermos2}
It is instructive to analyze the thermodynamic stability of $\caln=4$ SYM on
$S^3$ in the enthalpy representation ensemble, \ie when the
theory is subject to a fixed pressure $P$.
Here, the corresponding thermodynamic potential is the
enthalpy
\begin{equation}
H(Q_\alpha,S,P)\equiv \cale\cdot V + P\cdot V\equiv sT+\sum_\alpha \mu_\alpha\cdot Q_\alpha = 4 P\cdot V
\eqlabel{defh}
\end{equation}
where the last equality follows from the conformal symmetry of $\caln=4$ SYM. 
It is difficult to obtain $V(Q_\alpha,S,P)$ at
finite $K$; thus we work perturbatively in $\frac{K}{T^2}$, correspondingly
$S\gg 1$.
From \eqref{stuthermo} and \eqref{1stlaw1} we find
\begin{equation}
K=\frac{\sqrt{8P}\pi^{5/3}N_c^{1/3}}{S^{2/3}\prod_\alpha\left(1
+\frac{8\pi^2 Q_\alpha^2}{S^2}\right)^{1/6} }\cdot
\sum_{k=0}^\infty \bigg\{ v_k\cdot
\left({\pi N_c^2}\right)^{\frac{2k}{3}}
\cdot \left(\frac{1}{S^{\frac{2}{3}}}\right)^k\bigg\}\,,
\eqlabel{pstu2}
\end{equation}
with the first couple coefficients $v_k$ are given explicitly by 
\begin{equation}
v_1=\frac{-\frac 12-\sum_\alpha\frac{4\pi^2Q_\alpha^2}{3S^2}}
{\prod_\alpha\left(1+\frac{8\pi^2Q_\alpha^2}{S^2}\right)^{2/3}}\,,\qquad 
v_2=\frac{\frac38+\sum_\alpha\frac{2\pi^2Q_\alpha^2}{S^2}
+\sum_\alpha\frac{56\pi^4Q_\alpha^4}{9S^4}+\sum\sum_{\alpha> \beta}
\frac{16\pi^4Q_\alpha^2Q_\beta^2}{9S^4}}
{\prod_\alpha\left(1+\frac{8\pi^2Q_\alpha^2}{S^2}\right)^{4/3}}\,.
\eqlabel{defepsilon2}
\end{equation}
An equilibrium state is thermodynamically stable if the Hessian matrix of the second
derivatives of $H(Q_\alpha,S,P)$, \ie $\calh^H_{ij}\equiv \del^2E/\del y_i\del y_j$,
where $y_i=(Q_\alpha,S,P)$, $\alpha=\{3,2,1\}$, is positive definite
(convex) with respect to the extensive variables $(Q_\alpha,S)$, and
it has one negative direction associated with the intensive variable $P$
\cite{LandauLifshitzStatPhys}. The latter statement implies positive
compressibility (and thus mechanical stability) under the fixed pressure ---
in our case $\det \calh^H_{ij}<0$.

Recall that in the microcanonical ensemble the extensivity of the total
energy in volume in flat space implies the zero eigenvalue of the
Hessian $\calh^E_{ij}$ as $K=0$, see \eqref{eq4e}. There is
also a zero eigenvalue of the Hessian $\calh^H_{ij}$
in the flat space limit. Indeed, in the flat space, $K=0$,
conformality again implies extensivity at fixed pressure,
\begin{equation}
H(\lambda\cdot Q_\alpha\,,\lambda\cdot S,P)=\lambda\cdot H(Q_\alpha,S,P)\,,
\eqlabel{exth}
\end{equation}
implying that the Hessian $\calh^H_{ij}$ always has a null vector
$(Q_3,Q_2,Q_1,S,0)^T$. Evaluating the Hessian $\calh^H$ and its leading principle minors
\[
\left\{\Delta_1\equiv \frac{\del^2 H}{\del Q_3^2}\,,\qquad \cdots\,,\qquad 
\Delta_5\equiv \det \calh^H_{ij}\right\}\,,
\] we can express
the thermodynamic stability conditions in terms $\kappa_\alpha$ and $\frac{K}{(\pi T_0)^2}$:
\begin{equation}
\begin{split}
\Delta_{1}:\qquad
0<
\left(2-\kappa_{3}\right)
+
\frac{\kappa_{3}\left(10\kappa_{1}+10\kappa_{2}+7\kappa_{3}\right)}
{9\prod_{\alpha}(1+\kappa_{\alpha})}
\,\frac{K}{(\pi T_{0})^{2}}
+\mathcal{O}\!\left(\frac{K^{2}}{(\pi T_{0})^{4}}\right)\,,
\end{split}
\eqlabel{eq4a2}
\end{equation}
\begin{equation}
\begin{split}
&\Delta_{2}:\qquad
0<
\left(2-\kappa_{2}-\kappa_{3}\right)
-\frac{1}{9\prod_{\alpha}(1+\kappa_{\alpha})}
\biggl(
\kappa_{2}\kappa_{3}\left(20\kappa_{1}-\kappa_{2}-\kappa_{3}\right)
-\left(\kappa_{2}+\kappa_{3}\right)10\kappa_{1}
\\&-7\left(\kappa_{2}^{2}+\kappa_{3}^{2}\right)
-20\kappa_{2}\kappa_{3}
\biggr)
\frac{K}{(\pi T_{0})^{2}}
+\mathcal{O}\!\left(\frac{K^{2}}{(\pi T_{0})^{4}}\right)\,,
\end{split}
\eqlabel{eq4b2}
\end{equation}
\begin{equation}
\begin{split}
&\Delta_{3}:\qquad
0<
\left(2-\kappa_{1}-\kappa_{2}-\kappa_{3}+\kappa_{1}\kappa_{2}\kappa_{3}\right)
-\frac{1}{9\prod_{\alpha}(1+\kappa_{\alpha})}
\biggl(
9\kappa_{1}\kappa_{2}\kappa_{3}(\kappa_{1}+\kappa_{2}+\kappa_{3})
\\&-(\kappa_{1}+\kappa_{2}+\kappa_{3})(\kappa_{1}\kappa_{2}+\kappa_{1}\kappa_{3}+\kappa_{2}\kappa_{3})
-7(\kappa_{1}+\kappa_{2}+\kappa_{3})^{2}
-6(\kappa_{1}\kappa_{2}+\kappa_{1}\kappa_{3}+\kappa_{2}\kappa_{3})
\\&+63\kappa_{1}\kappa_{2}\kappa_{3}
\biggr)
\frac{K}{(\pi T_{0})^{2}}
+\mathcal{O}\!\left(\frac{K^{2}}{(\pi T_{0})^{4}}\right)\,,
\end{split}
\eqlabel{eq4cc}
\end{equation}
\begin{equation}
\begin{split}
\Delta_{4}:\qquad
0<
-\left(
2-\kappa_{1}-\kappa_{2}-\kappa_{3}
+\kappa_{1}\kappa_{2}\kappa_{3}
\right)K
+\mathcal{O}\!\left(\frac{K^{2}}{(\pi T_{0})^{4}}\right)\,,
\end{split}
\eqlabel{eq4d2}
\end{equation}
\begin{equation}
\begin{split}
&\Delta_{5}:\qquad
0>
-\left(2-\kappa_{1}-\kappa_{2}-\kappa_{3}
+\kappa_{1}\kappa_{2}\kappa_{3}\right)
-\frac{1}{\prod_{\alpha}(1+\kappa_{\alpha})}
\biggl(
2(\kappa_{1}+\kappa_{2}+\kappa_{3})^{2}
\\&-(\kappa_{1}\kappa_{2}+\kappa_{1}\kappa_{3}+\kappa_{2}\kappa_{3})
-2\kappa_{1}\kappa_{2}\kappa_{3}
(\kappa_{1}+\kappa_{2}+\kappa_{3}+3)
-3
\biggr)
\frac{K}{(\pi T_{0})^{2}}
+\mathcal{O}\!\left(\frac{K^{2}}{(\pi T_{0})^{4}}\right)\,.
\end{split}
\eqlabel{eq4e2}
\end{equation}
Notice that \eqref{eq4cc} and \eqref{eq4d2} are always mutually
incompatible: $U(1)^3$-charged $\caln=4$ SYM plasma is {\it always}
thermodynamically
unstable on $S^3$ (at least for small $K>0$) in the
enthalpy representation ensemble.

The flat space, \ie $K=0$, thermodynamic instability for a
{\it symmetric} state with $\kappa_\alpha\equiv \kappa$ is triggered
from violation of the $\Delta_2$ leading minor condition
\eqref{eq4b2}; this condition for the symmetric state becomes  
\begin{equation}
\begin{split}
&0<
2(1-\kappa)
+\frac{2\kappa^{2}(3-\kappa)}{(1+\kappa)^{3}}
\cdot\frac{K}{(\pi T_{0})^{2}}
+\frac{2\kappa^{2}(4\kappa^{2}-5\kappa-3)}{(1+\kappa)^{6}}
\cdot\frac{K^{2}}{(\pi T_{0})^{4}}
\\&-\frac{2\kappa^{2}(12\kappa^{3}-4\kappa^{2}-11\kappa-3)}{(1+\kappa)^{9}}
\cdot\frac{K^{3}}{(\pi T_{0})^{6}}
+\frac{2\kappa^{2}(32\kappa^{4}+12\kappa^{3}-24\kappa^{2}-17\kappa-3)}{(1+\kappa)^{12}}
\cdot\frac{K^{4}}{(\pi T_{0})^{8}}
\\&+\mathcal{O}\!\left(\frac{K^{5}}{(\pi T_{0})^{10}}\right).
\end{split}
\eqlabel{eq4c12}
\end{equation}
Eq.~\eqref{eq4c12} can be rephrased as a minimal curvature $K$ condition, necessary
to enforce the positivity $\Delta_2 >0$ for $\kappa>1$; remarkably, while
\eqref{eq4c12} differs from \eqref{eq4c1}, we find the constraint identical to \eqref{eq4c2}.

\subsection{Helicity $h=0$ sector fluctuations of the STU black hole}\label{stufluc}

In this section we discuss a decoupled subset of helicity $h=0$ sector fluctuations
of STU black holes in the symmetric state\footnote{With equal $U(1)$ charges, \ie
$\kappa_\alpha=\kappa$.}, triggering the low-temperature instability. This analysis
is an extension of \cite{Gladden:2024ssb} when the Schwarzschild horizon is
$S^3$ of curvature $K$.

The background geometry of the symmetric STU black hole is very simple: from \eqref{stubac},
\begin{equation}
\begin{split}
&\phi_i\equiv 0\,,\qquad a_\alpha\equiv a=\frac{\sqrt{2\kappa}\ \sqrt{\beta+K\kappa}}
{(1+\kappa)(1+\kappa r)}(1-r)\,,\qquad 
\beta=K+(\pi T_0)^2(1+\kappa)^3\,,\\
&c_1=\frac{(\pi T_0)\sqrt f}{(1+\kappa r)\sqrt r}\,,\qquad c_2=\frac{(\pi T_0)\sqrt{1+\kappa r}}{\sqrt r}\,,\qquad c_3=\frac{\sqrt{1+\kappa r}}{2 r\sqrt f}\,,\\
&f=(1+\kappa r)^3+\frac{Kr}{(\pi T_0)^2}-\frac{\beta r^2}{(\pi T_0)^2}\,.
\end{split}
\eqlabel{stufl1}
\end{equation}
Using the general expressions \eqref{w22}-\eqref{w0102}, we find that the master field
fluctuations defined as
\begin{equation}
\begin{split}
&{\hat F}^{(0)}_{(1,\alpha)}\equiv F^{(0)}_{(1,\alpha)}-\frac 13 \sum_{\beta=1}^3 F^{(0)}_{(1,\beta)}\,,\\
&{\hat F}^{(0)}_{(0,1)}\equiv -\frac 12  F^{(0)}_{(0,1)}-\frac {\sqrt3}{2} F^{(0)}_{(0,2)}\,,\qquad
{\hat F}^{(0)}_{(0,2)}\equiv -\frac 12  F^{(0)}_{(0,1)}+\frac {\sqrt3}{2} F^{(0)}_{(0,2)}\,,\qquad
{\hat F}^{(0)}_{(0,3)}\equiv F^{(0)}_{(0,1)}\,,
\end{split}
\eqlabel{deffh}
\end{equation}
decouple from the remaining fluctuations in the helicity-0 sector (see \eqref{gauge0m});
furthermore, as in \cite{Gladden:2024ssb}, there is a separate decoupling
for each index $\alpha$ of the sets $\{{\hat F}^{(0)}_{(1,\alpha)},{\hat F}^{(0)}_{(0,\alpha)}\}$,
leading to
\begin{equation}
\begin{split}
&0=\del^2_{rr}F_1 -\frac{c_3^2}{c_1^2}\ \del^2_{tt}F_1+
\left(\ln \frac{c_1c_2^3}{c_3}\right)'\ \del_r F_1+\left(\frac{4c_1'c_2'}{c_1c_2}-
\frac{c_3^2k^2}{c_2^2}\right)F_1-\frac{2 a'c_3 k}{\sqrt 3 c_1c_2} F_0\,,
\end{split}
\eqlabel{ffluc1}
\end{equation}
\begin{equation}
\begin{split}
&0=\del^2_{rr}F_0 -\frac{c_3^2}{c_1^2}\ \del^2_{tt}F_0+
\left(\ln \frac{c_1c_2^3}{c_3}\right)'\ \del_r F_0+\left(4c_3^2
-\frac{c_3^2k^2}{c_2^2}-\frac{(a')^2}{c_1^2}\right)F_0-\frac{\sqrt 3 a'c_3 k}{ c_1c_2} F_1\,,
\end{split}
\eqlabel{ffluc2}
\end{equation}
where $k$ is quantized as in \eqref{ks3}, and we introduced
\begin{equation}
\{{\hat F}^{(0)}_{(1,\alpha)},{\hat F}^{(0)}_{(0,\alpha)}\}\ \equiv \{F_1,F_0\}\,.
\eqlabel{deffs}
\end{equation}
Remarkably, there is no explicit $K$ dependence in \eqref{ffluc1} and \eqref{ffluc2} ---
the QNMs in this sector feel the horizon curvature only implicitly, through the background
geometry \eqref{stufl1}.

To determine the spectrum of QNMs, we solve \eqref{ffluc1} and \eqref{ffluc2},  
subject to infalling boundary condition at the horizon \cite{Son:2002sd}, \ie as $r\to 0$,
and the normalizability at the asymptotic $AdS_5$ boundary \cite{Kovtun:2005ev},
\ie as $r\to 0$. It is convenient to introduce
\begin{equation}
F_{1}(t,r)\equiv e^{-i 2\pi T \ww  t} (1-r)^{-i \ww/2}\ f_1(r)\,,\qquad
F_{0}(t,r)\equiv e^{-i 2\pi T \ww  t} (1-r)^{-i \ww/2}\ f_0(r)\,,
\eqlabel{bc}
\end{equation}
where $\ww=\frac{w}{2\pi T} $. Then the radial wavefunctions $f_1$ and $f_0$ must be regular
at the horizon, and vanish as $\calo(r)$ as $r\to 0$.

\bibliographystyle{JHEP}
\bibliography{n4s3}

\end{document}